\documentclass[showpacs,floatfix,eqsecnum,aps,nofootinbib]{revtex4}
\usepackage{graphics}

\begin{document}

\title{Spectator Effects in the Decay $B\rightarrow K\gamma\gamma$}
\author{A.~Yu.~Ignatiev}
\email{a.ignatiev@physics.unimelb.edu.au}
\author{G.~C.~Joshi}
\email{joshi@tauon.ph.unimelb.edu.au}
\author{B.~H.~J.~McKellar}
\email{b.mckellar@physics.unimelb.edu.au} \affiliation{
    ~{\em School of Physics, University of Melbourne,}\\
    {\em   Australia.}}

\def\be{\begin{equation}}
\def\ee{\end{equation}}
\def\bea{\begin{eqnarray}}
\def\eea{\end{eqnarray}}
\newcommand{\nn}{\nonumber \\}

\begin{abstract}
We report the results of the first computation related to the study of
the spectator effects in the rare decay mode $B\rightarrow K \gamma
\gamma $ within the framework of Standard Model. It is found that the
account of these effects results in the enhancement factor for the
short-distance reducible contribution to the branching ratio. 
\\
\end{abstract}

\maketitle

\begin{section}{Introduction}

The decays of  B mesons  provide us with a valuable tool for studying CP
violation, testing the Standard Model and looking for new physics. In
particular, it has become possible to study low-probability processes
such as various rare decay modes of B mesons. Some of these decay rates
have already been measured while other rare decays are expected to be
observed in the future.

One process of the latter category is the decay $B\rightarrow
K\gamma\gamma$. Previously, this decay mode and/or the related quark
process $b\rightarrow s\gamma\gamma$ have been studied in the papers
\cite{cho,lin,chang,reina,singer, ahmady1,ahmady2} where both
short-distance and long-distance contributions to the decay rate have
been considered. In Ref.~\cite{chang} it was pointed out that one of the
difficulties of the theoretical analysis is the account of spectator
effects in such decays. Spectator effects are those processes when one of
the photons is emitted by the u or d quark which are part of the decaying
B meson.

To the best of our knowledge, the contributions due to these effects
have not been estimated before and such an estimate is the purpose of
the present paper.
\end{section}
\begin{section}{Spectator effects}

 The effective Hamiltonian is \cite{lin}
\be
{\cal{H}}_{eff} = -\frac{G_F}{\sqrt{2}}V_{ts}^*V_{tb}\sum_i
                         C_i(\mu)O_i(\mu),
\ee with \bea O_1 & = & (\bar{s}_ic_j)_{V-A}(\bar{c}_jb_i)_{V-A}, \nn
O_2 & = & (\bar{s}_ic_i)_{V-A}(\bar{c}_jb_j)_{V-A}, \nn O_3 & = &
(\bar{s}_ib_i)_{V-A}\sum_q(\bar{q}_jq_j)_{V-A} \nn O_4 & = &
(\bar{s}_ib_j)_{V-A}\sum_q(\bar{q}_jq_i)_{V-A}, \nn O_5 & = &
(\bar{s}_ib_i)_{V-A}\sum_q(\bar{q}_jq_j)_{V+A}, \nn O_6 & = &
(\bar{s}_ib_j)_{V-A}\sum_q(\bar{q}_jq_i)_{V+A}, \nn O_7 & = &
\frac{e}{16\pi^2}\bar{s}_i\sigma^{\mu\nu}(m_sP_L + m_bP_R)b_iF_{\mu\nu},
\quad \quad \rm{and} \nn
O_8 & = & \frac{g}{16\pi^2}\bar{s}_i\sigma^{\mu\nu}(m_sP_L + m_bP_R)
           T_{ij}^ab_jG_{\mu\nu}^a.
\eea

We model the spectator effects by the loop diagram in Fig.~\ref{AD}.
\begin{figure}
\includegraphics{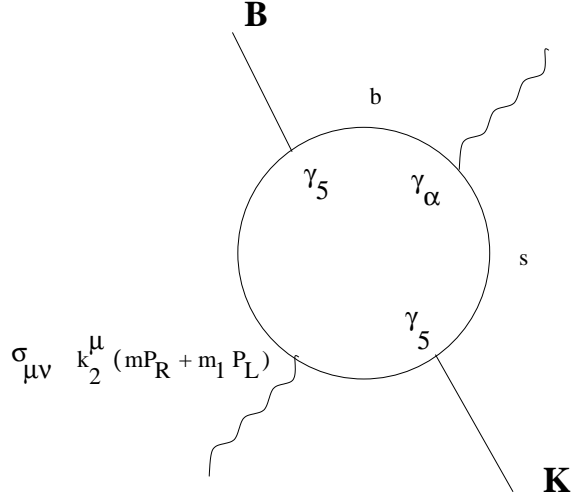}
\caption{\label{AD}Feynman diagram modelling the spectator effect in the
decay $B\rightarrow K\gamma\gamma$}.
\end{figure}
 The
corresponding amplitude is \bea \lefteqn{ M_{AD}= C\times}\nn & &  \int
d^4l\frac{Tr(\gamma_5 (\not l-\not k_1-\not k_2-\not q+m_2)\not
\epsilon_1(\not l-\not k_2-\not q)+m_2)\gamma_5((\not l-\not
k_2)+m_1)[-\not k_2, \not \epsilon_2](i/2)(m P_R+m_1 P_L)(\not l +m)
)}{(l^2-m^2)((l-k_1-k_2-q)^2-m_2^2)((l-k_2-q)^2-m_2^2)
((l-k_2)^2-m_1^2)}\nn & &  \eea plus Bose-exchanged part, where
\be
m\equiv m_b=4.8\; GeV,\hskip 1cm m_1\equiv m_s=0.15\; GeV,\hskip 1cm
m_2\equiv m_d=0.006\; GeV
 \ee
are the quark masses; $k_1,k_2$ and $q$ are the 4-momenta of the two
photons and the kaon, respectively.
 Neglecting the spectator effects the decay $B\rightarrow
K\gamma\gamma$ is similarly modelled by the diagrams in Fig.~\ref{SDa},
~\ref{SDb}.
\begin{figure}
\includegraphics{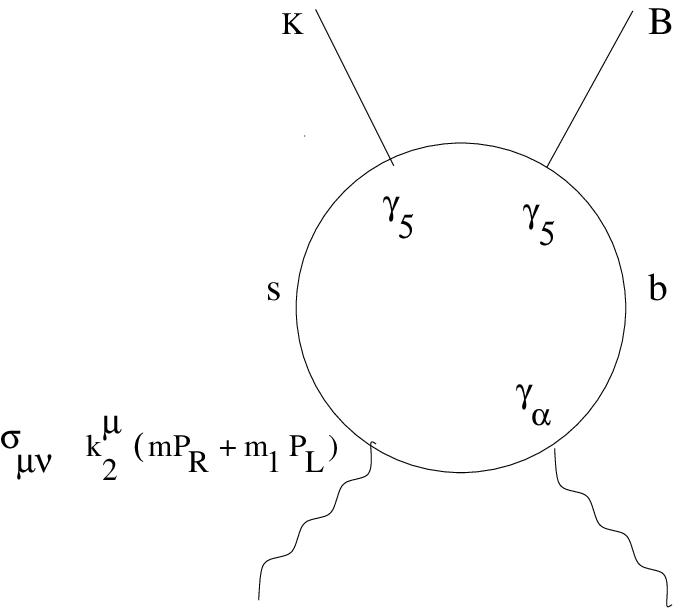}
\caption{\label{SDa}Feynman diagram modelling  the decay $B\rightarrow
K\gamma\gamma$ neglecting the spectator effects}.
\end{figure}
\begin{figure}
\includegraphics{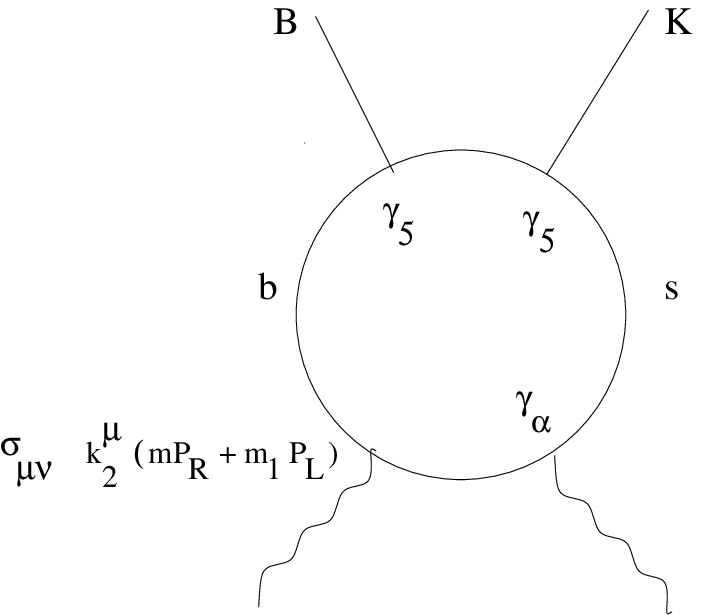}
\caption{\label{SDb}Feynman diagram modelling  the decay $B\rightarrow
K\gamma\gamma$ neglecting the spectator effects}.
\end{figure}
\begin{figure}
\includegraphics{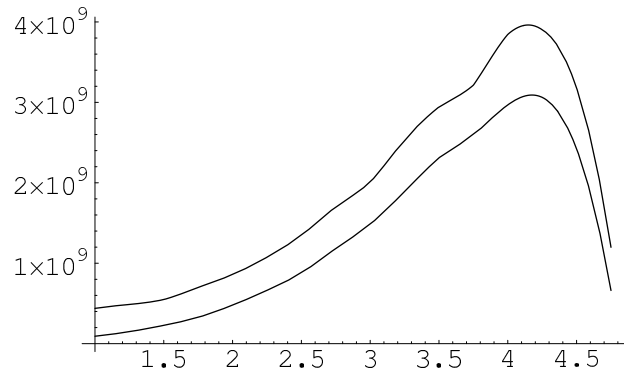}
\caption{\label{B0}The differential decay rates for the decay
$B^0\rightarrow K^0\gamma\gamma$ proceeding through the amplitude
$M_{SD}$ (bottom curve) and through the amplitude $ M_{SD}+M_{AD}$ (top
curve).}
\end{figure}
\begin{figure}
\includegraphics{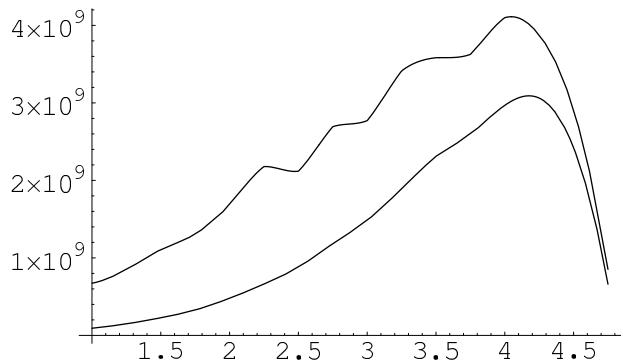}
\caption{\label{B+}The differential decay rates for the decay
$B^+\rightarrow K^+\gamma\gamma$ proceeding through the amplitude
$M_{SD}$ (bottom curve) and through the amplitude $ M_{SD}-2M_{AD}$ (top
curve). The factor (-2) comes from the ratio of the electric charges of
the spectator quark and non-spectator quark.}
\end{figure}
The amplitude is given by \be M_{SD}=M_{2a}+M_{2b}\ee \be M_{2a}=C\int
d^4l\frac{Tr(\not \epsilon_1(\not l+\not k_1+m)\gamma_5(\not l-\not
k_2-\not q+m_2)\gamma_5 (\not l-\not k_2+m_1)[-\not k_2, \not
\epsilon_2](i/2)(m P_R+m_1 P_L)(\not l +m)
)}{(l^2-m^2)((l-k_1-k_2-q)^2-m_2^2)((l +k_1)^2-m^2)
((l-k_2)^2-m_1^2)}\ee \be M_{2b}= C\int d^4l\frac{Tr(\not
\epsilon_1(\not l+\not k_1+m_1)[-\not k_2, \not \epsilon_2](i/2)(m
P_R+m_1 P_L)(\not l+\not k_1+\not k_2+m)\gamma_5(\not l-\not
q+m_2)\gamma_5(\not l+m_1))}{(l^2-m_1^2)((l+k_1+k_2)^2-m^2)((l
+q)^2-m_2^2) ((l+k_2)^2-m_1^2)} \ee plus Bose-exchanged parts.

All diagrams in Fig.~\ref{AD},~\ref{SDa} and ~\ref{SDb} are convergent.
To calculate the amplitudes we have used two software packages: FeynCalc
\cite{fc} and LoopTools \cite{lt}.

 To estimate the significance of
the spectator effects  we introduce the dimensionless ratio
\be
r=\frac{\Gamma_{SD+AD}}{\Gamma_{SD}}.
 \ee

 Here, $\Gamma_{SD+AD}$ is the contribution to the
the $B\rightarrow K\gamma\gamma$ decay rate including the spectator
effects while $\Gamma_{SD}$ is the the contribution neglecting the
spectator effects:
\be
\Gamma_i = \int \frac{1}{4m_B}|M_i|^2
\frac{d^3k_1}{2\omega_1(2\pi)^3}\frac{d^3k_2}{2\omega_2(2\pi)^3}\frac{d^3
q}{2\omega_K(2\pi)^3}(2\pi)^4 \delta (p_B-k_1-k_2-q). \ee

It was found in Ref.~\cite{cho} that contributions due to $\eta'$,
$\eta_c$ and $K^*$ were important. These contributions are not included
here. Therefore, it is more informative to consider the following ratios
with cuts: \be r_1=\frac{\Gamma_{SD+AD}(\sqrt{s_{\gamma
\gamma}}>m_{\eta_c}+2\Gamma_{\eta_c})}{\Gamma_{SD}(\sqrt{s_{\gamma
\gamma}}>m_{\eta_c}+2\Gamma_{\eta_c})} \ee

and
 \be
r_2=\frac{\Gamma_{SD+AD}(\sqrt{s_{\gamma
\gamma}}>m_{\eta_c}+2\Gamma_{\eta_c},
\sqrt{s_{K\gamma}}>m_{K^*}+2\Gamma_{K^*})}{\Gamma_{SD}(\sqrt{s_{\gamma
\gamma}}>m_{\eta_c}+2\Gamma_{\eta_c},\sqrt{s_{K\gamma}}>m_{K^*}+2\Gamma_{K^*})}.
\ee

 The calculation gives the results collected in
 Table 1 showing that the spectator effects contributions range from
 about 40 \% in the case of neutral B mesons to about 60 \% in the case
 of charged B mesons.
\begin{table}
\caption{Relative magnitudes of the spectator effects }
\begin{tabular}{|c|c|c|}\hline
   &$B^0\rightarrow K\gamma\gamma$ & $B^+\rightarrow K^+ \gamma\gamma$ \\
   \hline
  r & 1.38 & 1.65  \\\hline
  $r_1$ & 1.32 & 1.38 \\ \hline
  $r_2$ & 1.32 & 1.36 \\ \hline
\end{tabular}
\end{table}
Another way to represent the magnitude of the spectator effects is to
plot the differential decay rates $d\Gamma/d(\sqrt{s_{\gamma \gamma}})$
as functions of $\sqrt{s_{\gamma \gamma}}$ (in GeV) through the amplitude
$M_{SD}$ and through the combination of the amplitudes $M_{SD}$ and
$M_{AD}$, see Fig.~\ref{B0}  and \ref{B+} (because we are only interested
in the ratios of the decay rates, the units on the $y$-axis are
arbitrary). From Table 1 we see that that the cuts induce a larger shift
in $r$ in the case of the decay $B^+\rightarrow K^+ \gamma\gamma$ as
compared to the case of $B^0\rightarrow K\gamma\gamma$. This is
consistent with the fact that the gap between the two curves in Fig.~5 is
larger than the corresponding gap in Fig.~4.

\end{section}

\end{document}